\documentclass[aps,prl,twocolumn,floats]{revtex4}
\usepackage{epsfig}
\usepackage{amsmath}
\newcommand{\beq}{\begin{equation}}
\newcommand{\eeq}{\end{equation}}
\newcommand{\bea}{\begin{eqnarray}}
\newcommand{\eea}{\end{eqnarray}}
\newcommand{\ba}{\begin{array}}
\newcommand{\ea}{\end{array}}

\newcommand{\al}{\alpha}

\newcommand{\sech}{\mbox{\rm sech}}

\begin{document} 
\title{{\bf Light Bullet Modes in Self-Induced-Transparency Media with 
Refractive Index Modulation}} 
\author{Miriam Blaauboer,$^{{\rm a,b}}$ Gershon Kurizki,$^{{\rm a}}$ and 
Boris A. Malomed$^{{\rm b}}$}
\affiliation{
$^a$ Chemical Physics Department, Weizmann Institute of Science,\\
Rehovot 76100, Israel\\
$^b$ Department of Interdisciplinary Studies, Faculty of Engineering,\\
Tel Aviv University, Tel Aviv 69978, Israel}
\date{\today}

\begin{abstract}
We predict the existence of a new type of spatiotemporal solitons 
("light bullets") in two-dimensional self-induced-transparency 
media with refractive index modulation in the direction transverse 
to that of pulse propagation. These self-localized guided 
modes are found in an approximate analytical form, their 
existence and stability being confirmed by numerical simulations, 
and may have advantageous properties for signal transmission.
\end{abstract}


\maketitle

"Light bullets" are multi-dimensional solitons that are localized in both 
space and time. In the last decade they have been theoretically investigated
in various nonlinear optical 
media~\cite{silb90,haya93,malo97,miha98,He98,kher98,fran98,gott98,blaa00,blaa001}, 
and the first experimental
observation of a quasi two-dimensional (2D) bullet was recently reported~\cite
{liu99}. A promising candidate for the observation of fully 2D and 3D light
bullets is a self-induced-transparency (SIT) medium. SIT
involves undistorted and unattenuated propagation of an electromagnetic
pulse in a medium consisting of near-resonant two-level atoms, irrespective
of the carrier-frequency detuning from the resonance \cite{mcca67, maim90}. 
As early as 1984, simulations had shown self-focusing of 
spatiotemporal pulses in a SIT medium into a quasi-stable vibrating 
object~\cite{drum84}, thus hinting at the possible existence of ``bullets''. 
In recent works \cite{blaa00,blaa001}, we have predicted that both 
uniform 2D and 3D SIT media \cite{blaa00} and SIT media embedded in a Bragg 
grating \cite{blaa001} can support stable light bullets (LBs). Here we 
complement these investigations with the case of SIT media with refractive 
index (RI) modulation in the direction {\it transverse} to that of 
pulse propagation. We show that such a structure acts as a unique 
nonlinear waveguide, which transversely guides (confines) a new type of LBs, 
whereas other pulses are dispersed and diffracted \cite{ragh00}. Following 
the discovery of SIT soliton propagation in erbium-doped resonant fiber
waveguides \cite{naka92}, a demonstration of the existence of light 
bullet guided modes may lead to new possibilities for optical signal 
processing.

Our starting point is a 2D SIT medium with a spatially-varying refractive
index $n(z,x)$, which is described by the lossless Maxwell-Bloch equations 
\cite{newe92},
 
\begin{subequations}
\label{eq:SIT1}
\begin{eqnarray}
-i{\cal E}_{xx}+n^{2}\,{\cal E}_{\tau }+{\cal E}_{z}+i\,(1-n^{2})\,{\cal E}-
{\cal P} &=&0,  \label{eq:SIT11} \\
{\cal P}_{\tau }-{\cal E}W &=&0,  \label{eq:SIT12} \\
W_{\tau }+\frac{1}{2}({\cal E}^{\ast }{\cal P}+{\cal P}^{\ast }{\cal E})
&=&0.  \label{eq:SIT13}
\end{eqnarray}
\end{subequations}
Here ${\cal E}$ and ${\cal P}$ are the slowly varying amplitudes of the
electric field and polarization, $W$ is the inversion, $z$ and $x$ are the
longitudinal and transverse coordinates (in units of the
resonant-absorption length), and $\tau $ is time (in units of the
input pulse duration). The Fresnel number $F$, which governs the transverse
diffraction in 2D and 3D propagation, has been incorporated in $x$ and
the detuning $\Delta \Omega$ of the carrier frequency $\omega _{0}$ from the central
atomic-resonance frequency was absorbed in ${\cal E}$ and ${\cal P}$. 
They can be reintroduced into Eqs.~(\ref
{eq:SIT1}) by the transformation ${\cal E}(\tau ,z,x)\rightarrow 
{\cal E}(\tau ,z,x)\,\mbox{\rm exp}(-i\Delta \Omega\,
\tau )$, ${\cal P}(\tau ,z,x)\rightarrow {\cal P} 
(\tau ,z,x)\,\mbox{\rm exp}(-i\Delta \Omega\, \tau )$, and 
$x\rightarrow F^{-1/2}\,x$. We have neglected polarization dephasing 
and inversion decay by assuming 
pulse durations that are short on the time scale of these relaxation 
processes. Equations~(\ref{eq:SIT1}) are then compatible with the local 
constraint $|{\cal P}|^{2}+W^{2}=1$, which corresponds to Bloch-vector 
conservation \cite{newe92}. In the absence of the $x$-dependence and for $ 
n(z,x)=1$, Eq.~(\ref{eq:SIT11}) reduces to the sine-Gordon (SG) equation 
which has the soliton solution ${\cal E}(\tau ,z)=2\alpha \,\mbox{\rm sech} 
(\alpha \tau -z/\alpha +\Theta _{0})$, where $\alpha $ and $\Theta _{0}$ are 
real parameters. 

Our aim is now to investigate whether there exist stable light bullet
solutions of (\ref{eq:SIT1}) for a specified RI profile $n(x)$.
The physical idea behind this search is well-known: a specific transverse 
modulation of the refractive index can compensate for the transverse
diffraction of a specific pulse (described by the first term in (\ref{eq:SIT11})),
and thereby form a unique nonlinear waveguide confining that pulse 
whereas others may be diffracted. This phenomenon of guidance by 
transverse confinement has been demonstrated many times in waveguide 
theory and fiber optics \cite{levi80}, but as far as we know guidance of LBs 
has not yet been investigated. Since light bullets are not only spatially
but also temporarily localized, the possibility of "light bullet guided
modes" is both interesting from a fundamental point of view and may also
open new possibilities in signal transmission.

\begin{figure}[tbp]
%\centerline{\epsfig{figure=refr.eps,width=0.8\hsize}} 
\epsfig{figure=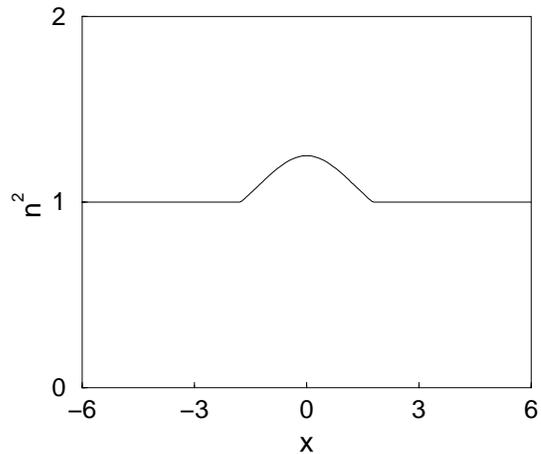,width=0.8\hsize}
\caption{The profile of the guiding structure corresponding to the modulated 
refractive index (\ref{eq:eps21}) with $\beta =0.5$.} 
\label{fig:diel} 
\end{figure}

We search for a stable solution of Eqs.~(\ref{eq:SIT1}) which 
(i) is simultaneously localized in time
and in the transverse direction, (ii) reduces to the SG soliton in 1D,
and (iii) whose transverse diffraction is compensated by a suitably chosen
RI modulation. The first criterion is satisfied for an electric field
amplitude of the variable-separated form ${\cal E}(\tau,z,x) = 
{\cal E}_{1}(\tau,z)\, {\cal E}_{2}(x)$. Using this form, we find 
an approximate analytical ansatz solution of (\ref{eq:SIT1})
which meets all three requirements:
\begin{subequations} 
\label{eq:parm2} 
\begin{eqnarray} 
{\cal E} &=&\pm \,2\,\alpha \,\mbox{\rm sech}\Theta (\tau ,z)\, 
\mbox{\rm sech}(\beta x)\, \exp(i\phi),  \label{eq:parm21} \\ 
{\cal P} &=&\pm \,2\,\mbox{\rm sech}\Theta (\tau ,z)\,\tanh \Theta (\tau 
,z)\,\mbox{\rm sech}(\beta x)\, \exp(i\phi),  \label{eq:parm22} \\ 
W &=&[1-4\,\mbox{\rm sech}^{2}\Theta (\tau ,z)\tanh ^{2}\Theta (\tau ,z) 
\mbox{\rm sech}^{2}(\beta x)]^{1/2},  \label{eq:parm23}
\end{eqnarray}
%\end{subequations} 
with
\bea 
\Theta (\tau,z) \equiv \alpha (\tau -z)- z/ \alpha + \Theta_{0},
\label{eq:Theta}
\eea
in the presence of the RI profile, see Fig.~\ref{fig:diel},
\bea
n^{2}(x) = \left\{ \begin{array}{ll}
1- \beta^{2} \left[ 1-2 \mbox{\rm sech}^{2}(\beta x)\right], 
& \mbox{\rm for}\, |x| \leq x_{\rm max} \\
1, & \mbox{\rm for}\, |x| \geq x_{\rm max} 
\end{array} \right.
\label{eq:eps21}
\eea
\end{subequations}
Here $x_{\rm max} \equiv \mbox{\rm arcsech} (1/ \sqrt{2})/\beta$, and $\alpha $, 
$\beta$, $\phi$ and $\Theta_{0}$ are real constants. The ansatz~(\ref{eq:parm2}) 
satisfies Eq.~(\ref{eq:SIT11}) exactly, while Eqs.~(\ref{eq:SIT12}) 
and (\ref{eq:SIT13}) are satisfied in the limit $\beta |x|\ll 1$. 
Numerical results displayed below verify that 
Eqs.~(\ref{eq:parm2}) indeed approximate a solution to Eqs.~(\ref{eq:SIT1}) to a 
high accuracy. The field (\ref{eq:parm21}) represents a LB traveling in the 
$z$-direction with a velocity $v=\alpha ^{2}/(\alpha ^{2}+1)$ and reduces
to the SG soliton in the limit $\beta \rightarrow 0$, when the
transverse guidance disappears. Hence this approximate 
"light bullet" solution is in fact a {\it nonlinear multi-dimensional  
guided mode}. Just as the 1D SG soliton it propagates undistorted and
unattenuated and satisfies the SIT area-theorem \cite{mcca67} $\int_{-\infty}^{\infty}
d\tau | {\cal E}(\tau,z,x=0) | = \pm 2\pi$. The LB mode (\ref{eq:parm2}) 
is unrelated to the LB found in Ref.~\cite{blaa00} in a uniform SIT medium;
in fact, it can be proven that light bullets of variable separated form
as in (\ref{eq:parm2}) do not exist in a uniform SIT medium \cite{blaa00}, 
nor in a SIT medium embedded in a Bragg grating in which the refractive 
index varies in the longitudinal direction \cite{blaa001}. Eq.~(\ref{eq:parm2})
represents a new type of nonlinear guided mode. 

\begin{figure} 
\centerline{\epsfig{figure=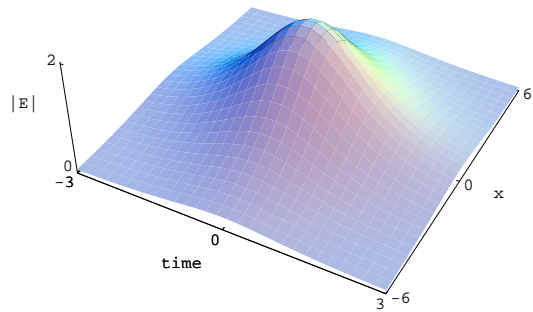,width=0.8\hsize}} 
\caption{The electric field inside the 2D ``light bullet'', $|{\cal E}|$, in 
a SIT medium with a transversely-modulated refractive index as per Eq.~(\ref{eq:eps21}), 
vs. time $\tau $ and transverse coordinate $x$ after 
having propagated the distance $z=1000$. Parameters used correspond to 
$\alpha=1$, $\beta =0.5$, $\phi =0$ and 
$\Theta_{0}=1000$.} 
\label{fig:field} 
\end{figure}

In order to test the accuracy and stability of the ansatz (\ref{eq:parm2}),
we have performed numerical simulations. 
Figure~\ref{fig:field} shows the electric field at $z=1000$, 
generated by direct simulation \cite{drum83} of the 2D SIT equations 
(\ref{eq:SIT1}) 
using Eqs. (\ref{eq:parm2}) as an initial configuration at $ 
z=0$. To a very good accuracy (with a deviation $<1\%$), the result of the 
evolution over this propagation distance, which is much larger than 
the corresponding diffraction length, still coincides with 
the initial field configuration.
The corresponding polarization distribution (not shown) has the shape of a 
double-peaked bullet in $\tau $ and decays in a similar way as the electric 
field, which is a characteristic property of SIT \cite{newe92}. Also the 
inversion is localized in both $\tau $ and $x$, taking at infinity the value  
$-1$, corresponding to the atoms in the ground state. We have checked 
that no instability of the LB solution sets in as long as the simulations were run, 
up to $z\sim 10^4$. 
In addition, we have gathered numerical evidence that the modulation 
(\ref{eq:eps21})
uniquely confines the LB mode (\ref{eq:parm2}), whereas other pulses experience 
transverse diffraction. Starting e.g. with the light bullet ansatz from Ref.~\cite{blaa00} 
for a uniform SIT medium, we found that they diffract after propagating
a distance $x\sim10$ \cite{unif}. An accurate determination of the degree of 
uniqueness with which a 
particular profile corresponds to a particular LB guided mode 
requires further investigations though. Similarly, it is interesting to
investigate how different RI modulations 
may guide other types of LB modes \cite{other2D,3D}.

Experimentally, nanolithography techniques allow for fabrication of 
dielectric structures with layer thicknesses $\sim $ a few atomic layers 
\cite{khit99}, and the study of light-matter interactions in such structures
has developed into a vast research area \cite{dowl00}. In particular, erbium-doped
resonant fiber waveguides, which were developed about a decade ago and
in which SIT has been observed \cite{naka92}, form interesting candidates for opening
new fields in optical communications. Generation of light
bullet modes in transversely modulated SIT structures presents a new experimental 
challenge in this field. We briefly discuss here experimental 
conditions under which the guided LB modes predicted above may be observed. 
A suitable SIT medium consists of rare-earth ions embedded in  
a dielectric structure, with a typical 
density $\sim 10^{16}$ cm$^{-3}$ \cite{khit99} and radiative relaxation 
time (at cryogenic temperatures) $\sim 100$ ns \cite{grei99}. 
The incident optical pulse, of duration $\tau _{p}<0.1$ ns,  
should have uniform 
transverse intensity and, for the transverse diffraction to be strong 
enough, one needs $\alpha _{{\rm eff}}d^{2}/\lambda _{0}<1$ \cite{slus74}, where 
$\alpha _{{\rm eff}}$, $\lambda _{0}$ and $d$ are the inverse resonant-absorption 
length, carrier wavelength, and pulse diameter resp. For 
$\alpha_{{\rm eff}}\sim 10^{4}$\thinspace\ m$^{-1}$ and $\lambda _{0}\sim 10^{-4}$ 
\thinspace\ m, one thus requires $d<10^{-4}$ m, which implies that the 
transverse medium size $L_{x}\sim \,1-10\,\mu $m. The parameter $\beta$
in Eq.~(\ref{eq:parm2}) is determined by the transverse component of
the wavevector $\kappa_{x}$ and we find $\kappa_{x} L_{x} \sim 0.01-0.1$,
which is in accordance with the physically relevant regime of $n(x) \geq 1$. 
The parameter $\alpha$, which determines the amplitude of the 
bullet and its localization in $\tau $ and $z$, is given by 
$\alpha =\sqrt{v\alpha _{{\rm eff}}\tau _{p}(1-v\alpha _{{\rm eff}}\tau _{p})}$, 
and can thus be controlled by the incident pulse 
duration and velocity. For rare-earth media, one typically has $\alpha \sim 
0.1-10$. Embedding this medium into a set of 10-100 transverse 
layers, each having a thickness on the order of the wavelength, and RI 
varying by $\sim $\thinspace\ 1\% from layer to layer \cite{levi80}, completes the 
setup for the observation of the 2D guided-LB modes. 
These are then expected to be localized on a time scale $\sim 100$ ps and 
transverse length scale $\sim 1\,\mu $m. 

To conclude, we have proposed and studied the novel regime of light bullets in 
SIT\ media 
embedded in transversely modulated structures. Such a structure is 
predicted to allow one to produce nonlinear guided light bullet modes 
of a specific and fully controllable shape, while 
causing other pulses to disperse via transverse diffraction.
Because of their simultaneous spatial and temporal localization, these 
LB modes may open new possibilities of signal transmission in nonlinear waveguides
and we hope they will trigger new experiments. 

B.A.M. gratefully acknowledges a valuable discussion with P.D. Drummond. 
This work was supported by the Israel Council for 
Higher Education, Israel Science Foundation, Minerva and the
Binational (US-Israel) Science Foundation through grant No. 1999459.


\begin{thebibliography}{99}
\bibitem{unif}
These bullets can be guided in the presence of 
a different refractive index 
modulation, given by $n(x) = 1 - \frac{1}{2} C^2 (-1 + \frac{1}{2} 
\sech^2 \Theta_{1} + \frac{1}{2} \sech^2 \Theta_{2} + \tanh \Theta_{1}
\tanh \Theta_{2})$, with $C$, $\Theta_{1}$ and $\Theta_{2}$ as defined
in Ref.~\cite{blaa00}.
\end{thebibliography}
\end{document}